\documentclass[aps,prb,twocolumn,superscriptaddress,showpacs,preprintnumbers,amsmath,amssymb]{revtex4-2}
\usepackage{graphicx}
\usepackage{dcolumn}
\usepackage{bm}
\usepackage{txfonts}
\usepackage{float}
\usepackage{latexsym}

\usepackage{color}
\usepackage[normalem]{ulem}

\begin{document}


\title{Revised phase diagram of the high-$T_c$ cuprate superconductor Pb-doped Bi$_2$Sr$_2$CaCu$_2$O$_{8+\delta}$ revealed by anisotropic transport measurements} 

\author{Keiichi Harada}
\affiliation{Graduate School of Science and Technology, Hirosaki University, Hirosaki, Aomori, 036-8561 Japan}
	\author{Yuki Teramoto}
\affiliation{Graduate School of Science and Technology, Hirosaki University, Hirosaki, Aomori, 036-8561 Japan}
	\author{Tomohiro Usui}
\affiliation{Graduate School of Science and Technology, Hirosaki University, Hirosaki, Aomori, 036-8561 Japan}
	\author{Kenji Itaka}
 \affiliation{Institute of Regional Innovation, Hirosaki University, Aomori, Aomori, 036-8561 Japan}
        \author{Takenori Fujii}
\affiliation{Cryogenic Research Center, University of Tokyo, Bunkyo, Tokyo 113-0032, Japan}
        \author{Takashi Noji}
 \affiliation{Graduate School of Engineering, Tohoku University, Sendai 980-8579, Japan}
	\author{Haruka Taniguchi}
\affiliation{Graduate School of Engineering, Iwate University, Morioka 020-8551, Japan}
	\author{Michiaki Matsukawa}
\affiliation{Graduate School of Engineering, Iwate University, Morioka 020-8551, Japan}
	\author{Hajime Ishikawa}
\affiliation{Institute for Solid State Physics, University of Tokyo, Kashiwa, Chiba 277-8581, Japan}
        \author{Koichi Kindo}
\affiliation{Institute for Solid State Physics, University of Tokyo, Kashiwa, Chiba 277-8581, Japan}
        \author{Daniel S. Dessau}
 \affiliation{Department of Physics, University of Colorado at Boulder, Boulder, CO 80309, USA}

	\author{Takao Watanabe}
\email{twatana@hirosaki-u.ac.jp}
\affiliation{Graduate School of Science and Technology, Hirosaki University, Hirosaki, Aomori, 036-8561 Japan}





\date{\today}

\begin{abstract}
Although phase diagrams can be leveraged to investigate high transition temperature (high-$T_c$) superconductivity, the issue has not been discussed thoroughly. In this study, we elucidate the phase diagram of the overdoped side of high-$T_c$ cuprates via systematic anisotropic transport measurements for Pb-doped Bi-2212 single crystals. We demonstrate that the characteristic temperatures of the ``weak'' pseudogap opening and electronic coherence cross each other at a critical doping level, while those of the ``strong'' pseudogap merges into that of superconducting fluctuations above the critical doping level. Our results indicate the importance of Mottness in high-$T_c$ superconductivity.

\end{abstract}

\maketitle

\section{INTRODUCTION}
\begin{figure}[t]
		\begin{center}
			\includegraphics[width=85mm]{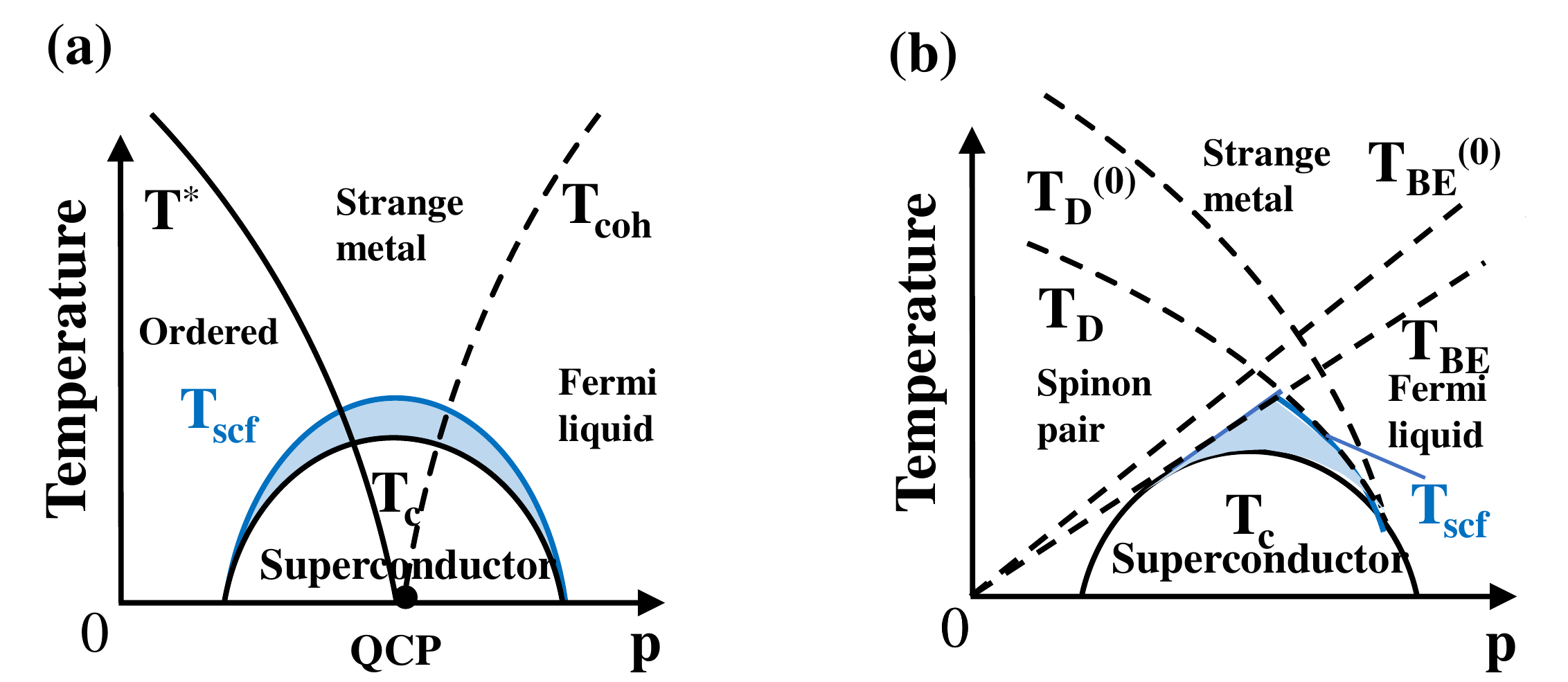}
			\caption{\label{fig1}(Color online) Schematic phase diagrams for (a) a QCP model \cite{Varma99,Sachdev10} and (b) an RVB model considering gauge field fluctuations \cite{Lee06,Ogata08,Lee92}. In each figure, the black bold lines denote the true phase transition temperatures, whereas the black dashed lines denote the crossover temperatures. The blue shaded area denotes the superconducting fluctuation regime. Here, the original phase transition temperatures of the RVB model are represented by $T_{scf}$, considering the strong two-dimensionality of cuprates.}   
		\end{center}
	\end{figure}

Superconductivity is an instability of the normal metallic state. Therefore, understanding the normal state from which superconductivity emerges is crucial to probe the underlying mechanism of superconductivity \cite{Norman05}. The features of the normal state can be evaluated from the doping ($p$)-temperature ($T$) phase diagram. Two types of theoretical phase diagrams have been proposed for cuprates. The first is the quantum critical point (QCP) model \cite{Varma99,Sachdev10}, in which $T^{*}$ (characteristic temperatures of pseudogap opening) and $T_{coh}$ (crossover temperatures from the ``strange metal'' state at higher temperatures to the Fermi-liquid-like state at lower temperatures) vanish at the QCP (Fig. \ref{fig1}(a)). Moreover, the fluctuations associated with the quantum critical phase transition mediate Cooper pairing. The second is the resonating valence bond (RVB) model \cite{Lee06,Ogata08,Lee92}. Here, we evaluate the RVB model by considering the fluctuations of the gauge fields to which spinons and holons are strongly coupled, since these fluctuations affect charge dynamics such as transport properties \cite{Lee06,Ogata08,Lee92}. In this model, spin gap opening temperature, $T_{D}$ (or its mean-field solution, $T_{D}^{(0)}$) and Bose condensation temperature for holons, $T_{BE}$ (or its mean-field solution, $T_{BE}^{(0)}$), which corresponds to $T_{coh}$  cross each other at a finite temperature near optimal doping (Fig. \ref{fig1}(b)). Here, large spin correlations arising from Mottness serve as the source for Cooper pairing. However, the level of accuracy of these two-phase diagrams has been debated upon in literature \cite{Keimer15}.

The positional relationship between the pseudogap (or spin gap) opening temperatures and $T_{coh}$ is unclear. In contrast to the pseudogap opening temperature \cite{Timusk99,Norman05,Hufner08,Kordyuk15,Vishik18}, the behavior of $T_{coh}$ has rarely been reported \cite{Kaminski03,Hussey11,Hussey13, Chatterjee11}, and a general consensus has not been obtained yet \cite{Keimer15}. This can be attributed to the limited number of materials that can be examined. Pb-doped Bi$_2$Sr$_2$CaCu$_2$O$_{8+\delta}$ (Bi-2212) (hereafter we denote it as Bi(Pb)-2212) is a suitable choice for such an investigation because the doping levels can be controlled from an approximately optimal value to beyond the critical doping level $p$ = 0.19 via oxygen annealing \cite{Tallon20,Chen19}. 

\begin{figure}[t]
		\begin{center}
			\includegraphics[width=85mm]{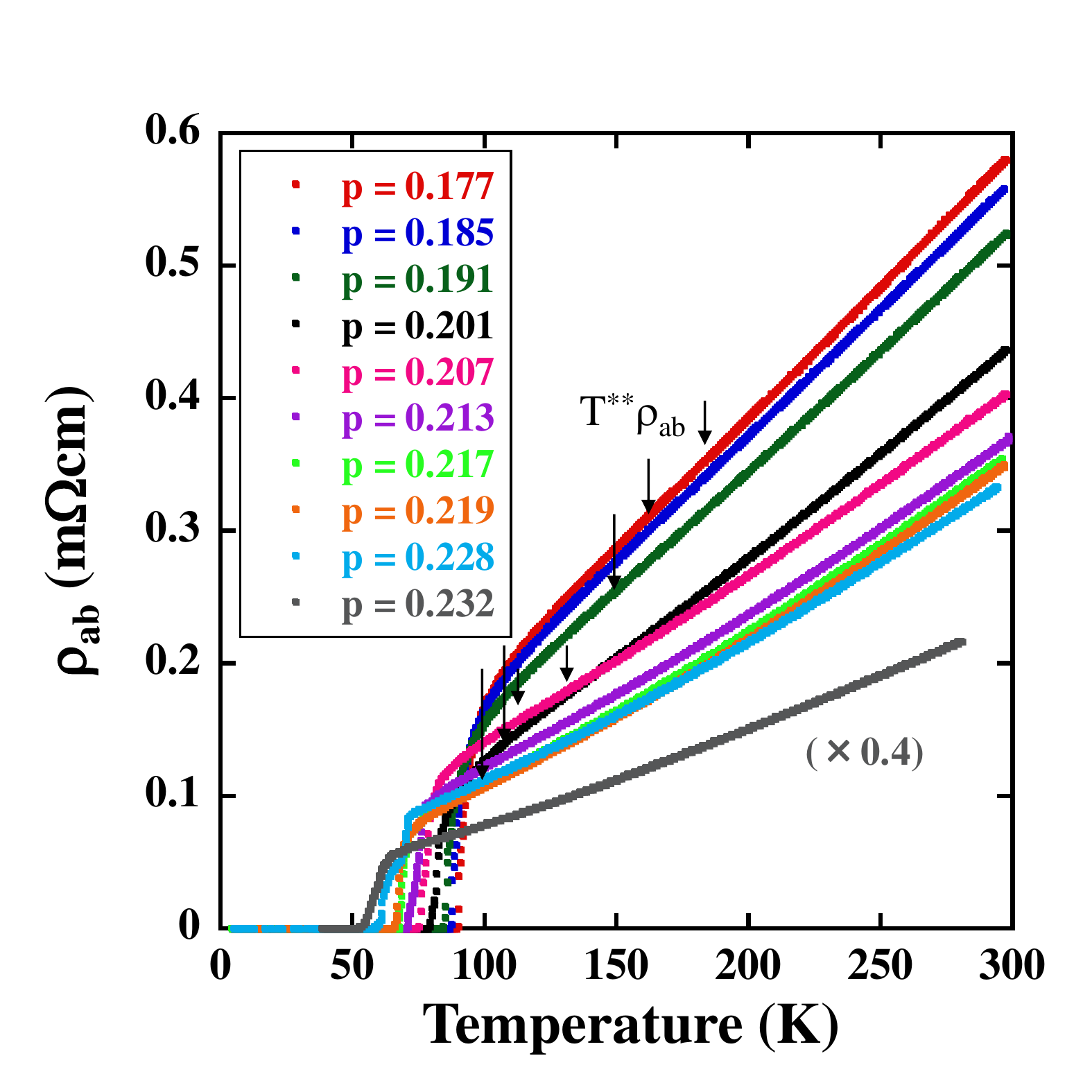}
			\caption{\label{fig2}(Color online) In-plane resistivity $\rho_{ab}(T)$ for Bi$_{1.6}$Pb$_{0.4}$Sr$_2$CaCu$_2$O$_{8+\delta}$ single crystals with various $p$. The most overdoped sample with $p$ = 0.232 is pristine Bi$_{2}$Sr$_2$CaCu$_2$O$_{8+\delta}$ \cite{Usui14}. The temperatures $T^{**}_{\rho_{ab}}$ below which $\rho_{ab}(T)$ decreases rapidly are indicated by arrows.}   
		\end{center}
	\end{figure}

\begin{figure*}[t]
		\begin{center}
			\includegraphics[width=180mm]{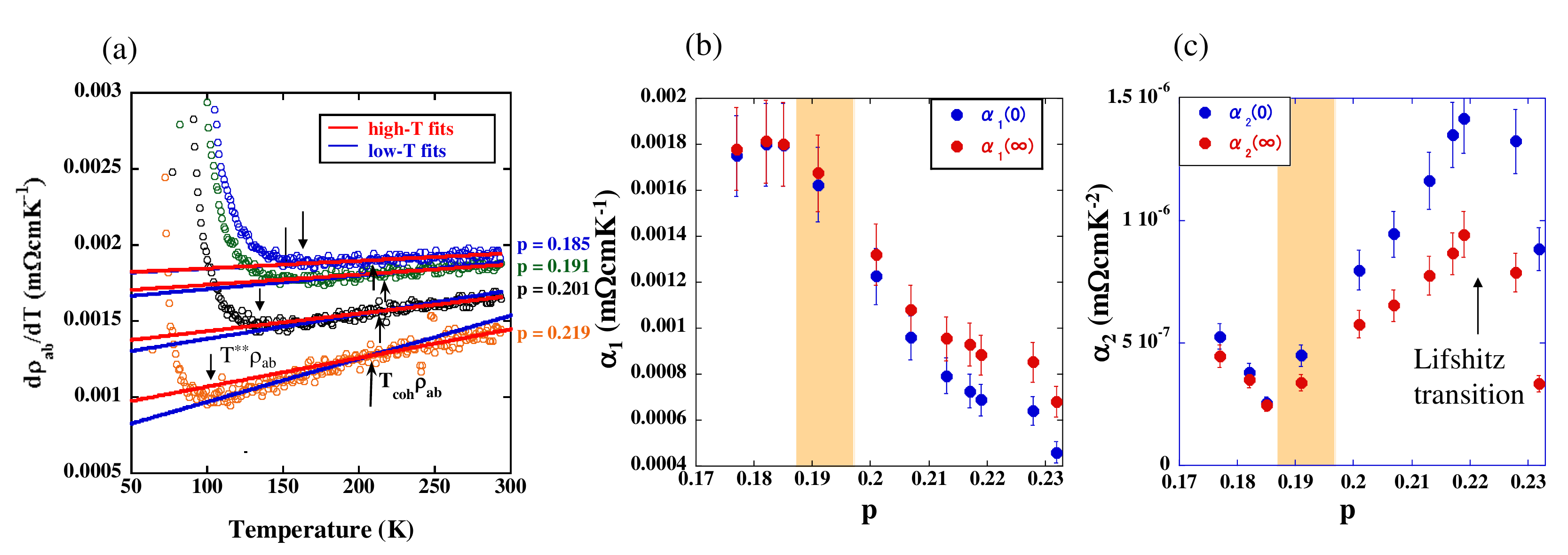}
			\caption{\label{fig3}(Color online) (a) Temperature derivative of $\rho_{ab}$ vs. temperature for several $p$. The temperatures $T^{**}_{\rho_{ab}}$ at which $d\rho_{ab}/dT$ is minimum and $T_{coh}{\rho_{ab}}$ at which high-temperature and low-temperature linear fits intersect are indicated by arrows. (b) The coefficient of the $T$-linear term in $\rho_{ab} (T)$, $\alpha_{1}$, vs. $p$. $\alpha_{1} (\infty)$ and $\alpha_{1} (0)$ is obtained by the linear fits for $d\rho_{ab}/dT$ at high and low temperatures, respectively. (c) The coefficient of the $T^2$ term in $\rho_{ab} (T)$, $\alpha_{2}$, vs. $p$. $\alpha_{2} (\infty)$ and $\alpha_{2} (0)$ is obtained by the linear fits for $d\rho_{ab}/dT$ at high and low temperatures, respectively.}
		\end{center}
	\end{figure*}

To this end, we first measured the anisotropic transport properties of Bi(Pb)-2212, and estimated the characteristic temperatures at which the typical temperature evolution occurs. $\rho_{c} (T)$ aids in identifying opening energy gaps, since it directly probes the electronic DOS around the Fermi level, reflecting the tunneling nature between CuO$_2$ planes in high-$T_c$ cuprates \cite{Watanabe00,Shibauchi01,Usui14}. Furthermore, $\rho_{ab} (T)$ is less sensitive to opening energy gaps \cite{Watanabe97,Naqib16,Usui14}, but sensitive to electronic coherence \cite{Hussey11,Hussey13} and superconducting fluctuations \cite{Usui14}. Systematic measurements of both $\rho_{ab} (T)$ and $\rho_{c} (T)$ for the same set of samples aid in obtaining the true phase diagram of overdoped Bi(Pb)-2212. The obtained phase diagram shows that i) two types (``weak'' and ``strong'') of pseudogaps exist, ii) they terminate at different $p^\prime$s, and iii) the characteristic temperatures of the ``weak'' pseudogap opening and $T_{coh}$ intersect. Finally, we discuss the implications of the obtained phase diagram.

\section{EXPERIMENT}
Single crystals of Bi$_{1.6}$Pb$_{0.4}$Sr$_2$CaCu$_2$O$_{8+\delta}$ (nominal composition of Bi$_{1.6}$Pb$_{0.6}$Sr$_2$CaCu$_2$O$_{8+\delta}$) were grown in air using the traveling solvent floating zone method. The crystals were annealed by varying the oxygen partial pressure, $P_{O2}$, (2 Pa $\le$ $P_{O2}$ $\le$ 400 atm), at 400--600 $^\circ$C for sufficient durations to homogeneously control the corresponding doping levels (for detailed annealing conditions, see Supplemental Materials \cite{Supplemental}). Moreover, previously reported data for excessively overdoped ($p$ = 0.232) pristine samples \cite{Usui14} were included in the analysis. We determined $T_c$ at the onset of zero resistivity. The doping level ($p$) was obtained using the empirical relation \cite{Obertelli92}, with maximum $T_c$ = 91.7 K and 91.0 K for Pb-doped single crystals and pristine samples, respectively.

$\rho_{ab} (T)$ and $\rho_{c} (T)$ were measured using the DC four-terminal method. In addition, to estimate the characteristic temperature for the superconducting fluctuation, $T_{scf}$, below which the superconducting fluctuation effect become appreciable, $\rho_{ab} (T)$ was measured using a physical property measurement system (Quantum Design) under various magnetic fields up to 9 T parallel to the $c$-axis. Moreover, to examine the temperature dependence of the Hall mobility, $\mu (T)$ (= $R_{H} (T)/\rho_{ab} (T)$), the Hall coefficient $R_{H} (T)$ was measured via the five-terminal method. Here, $R_{H}$ (= $\rho_{yx} (B)/B$) was obtained by averaging the difference of the Hall resistivity, $\rho_{yx}$, at positive and negative fields $B$, i.e., $\rho_{yx}(B) = (\rho_{yx}(+B) - \rho_{yx}(-B))/2$, which can eliminate the magnetoresistance (MR) component due to the misalignment of contacts.

\section{RESULTS}
Figure \ref{fig2} shows the temperature dependence of $\rho_{ab} (T)$ for Bi$_{1.6}$Pb$_{0.4}$Sr$_2$CaCu$_2$O$_{8+\delta}$ single crystals with various $p$. $T_c$ decreases systematically from 89.3 K to 65.0 K (52.0 K for the pristine sample) with increasing oxygen content, indicating that the measured samples are in the overdoped region. In the slightly overdoped region (0.177 $\le$ $p$ $\le$ 0.191), $\rho_{ab} (T)$ is approximately linear in $T$, which is consistent with a previous report \cite{Watanabe97}. However, by further increasing the doping level, $\rho_{ab} (T)$ exhibits a typical upward curvature.

To investigate the temperature dependence of $\rho_{ab} (T)$ in detail, its derivative with respect to temperature for several values of $p$ is plotted as a function of the temperature in Fig. \ref{fig3} (a). At higher temperatures, the derivative is approximately linear in $T$, whereas below the temperature region 180--100 K, it exhibits a steep upward curvature, reflecting a steep decrease in $\rho_{ab} (T)$ upon cooling. Here, we define the characteristic temperature at which the temperature derivative of $\rho_{ab} (T)$ is minimized as $T^{**}_{\rho_{ab}}$. $T^{**}_{\rho_{ab}}$ is estimated as 160, 150, 130, and 102 K for $p$ = 0.185, 0.191, 0.201, and 0.219, respectively. $T^{**}_{\rho_{ab}}$ may correspond to $T^{*}_{\rho_{ab}}$, which was previously defined for underdoped Bi-2212 \cite{Watanabe97,Usui14}.

The high-temperature $T$-linear behavior of the temperature derivative of $\rho_{ab} (T)$ indicates that $\rho_{ab} (T)$ can be expressed as \cite{Cooper09,Hussey11},

\begin{equation}
\rho_{ab} (T) = \alpha_{0} + \alpha_{1}T + \alpha_{2}T^2, 
\end{equation}
where $\alpha_{0}$ is the residual resistivity, $\alpha_{1}$ is the coefficient of the non-Fermi-liquid $T$-linear term, and $\alpha_{2}$ is the coefficient of the Fermi-liquid $T^2$ term. The slope and $y$-intercept of $d\rho_{ab}/{dT}$ represent 2$\alpha_{2}$ and $\alpha_{1}$, respectively. However, a closer look at the data reveals that the slope below 200 K is slightly larger than that above 200 K. Thus, the coefficients are obtained below and above 200 K separately by the linear fits (Fig. \ref{fig3} (a); for more information on the linear fits, see Supplemental Materials \cite{Supplemental}), and are denoted as $\alpha_{1} (0)$, $\alpha_{2} (0)$, and $\alpha_{1} (\infty)$, $\alpha_{2} (\infty)$, respectively. The rise (fall) in $\alpha_{2}$($\alpha_{1}$) upon cooling implies that the electronic state approximates that of Fermi liquids in the lower temperature region. Based on this, $T_{coh}{\rho_{ab}}$ is defined by the temperature at which both linearly fitted lines intersect. As expected, $T_{coh}{\rho_{ab}}$ is estimated to be approximately 200 K for all doping levels (Fig. \ref{fig3} (a)). 

However, it should be noted that the change above and below $T_{coh}{\rho_{ab}}$ is insignificant for $p$ $<$ 0.19 (see Supplemental Materials \cite{Supplemental}). The reason will be discussed later in the final paragraph of section IV. Furthermore, the slope of $d\rho_{ab}/{dT}$ is finite for all $p$, indicating that $\rho_{ab} (T)$ is not strictly $T$-linear even in the higher temperature region. This result is different from the typical behaviors for the ``strange metal'' state \cite{Varma99,Sachdev10}, and from previous reports on Bi-2212 \cite{Kaminski03}, La$_{2-x}$Sr$_x$CuO$_4$ (LSCO) \cite{Hussey11}, and Tl$_2$Ba$_2$CuO$_{6+\delta}$ (Tl-2201) \cite{Hussey13}. However, this result agrees with Bi$_2$Sr$_{2-x}$La$_x$CuO$_6$ (BSLCO) \cite{Ando99}, in which a finite slope in $d\rho_{ab}/{dT}$ has been observed for La content $x$ = 0.24 and 0.30.

Figures \ref{fig3}(b) and \ref{fig3} (c) depict the doping level dependence of $\alpha_{1}$ and $\alpha_{2}$, respectively. Upon increasing the doping level to above 0.19, $\alpha_{1}$($\alpha_{2}$) decreases (increases), implying that the system acquires a Fermi-liquid-like nature over this doping level range. This result is consistent with recent ARPES measurements of Bi(Pb)-2212, which reveal that with the increasing doping level across $p$ = 0.19, the incoherent spectral function abruptly reconstructs into a coherent one near the Brillouin zone boundary \cite{Chen19}. The decrease in $\alpha_{1}(0)$ for $p$ $>$ 0.19 is consistent with previous reports for LSCO \cite{Cooper09,Hussey11} and Tl-2201 \cite{Hussey13}. However, with a further increase in the doping level, $\alpha_{2}$ exhibits a peak at approximately $p$ = 0.22. This may be related to the reported Lifshitz transition at $p$ = 0.22, where the Fermi surface originating from the antibonding states changes its topology from an open hole-like state to a closed electron-like state \cite{Kaminski06,Benhabib15}. The van Hove singularity then crosses the Fermi level, which causes a peak in the specific heat coefficient, $\gamma$. The observed peak in $\alpha_{2}$ may be attributed to the empirical fact that the Kadowaki--Woods ratio, $\alpha_{2}/\gamma^2$, is maintained constant within the same series of compounds \cite{Kadowaki86,Hussey05}. In addition, we find that the difference between $\alpha_{1} (0)$($\alpha_{2} (0)$) and $\alpha_{1} (\infty)$($\alpha_{2} (\infty)$) is expanded continuously for $p$ $\ge$ 0.19. This implies that the system becomes more Fermi-liquid-like below $T_{coh}{\rho_{ab}}$ with an increase in the doping level.

	\begin{figure}[t]
		\begin{center}
			\includegraphics[width=85mm]{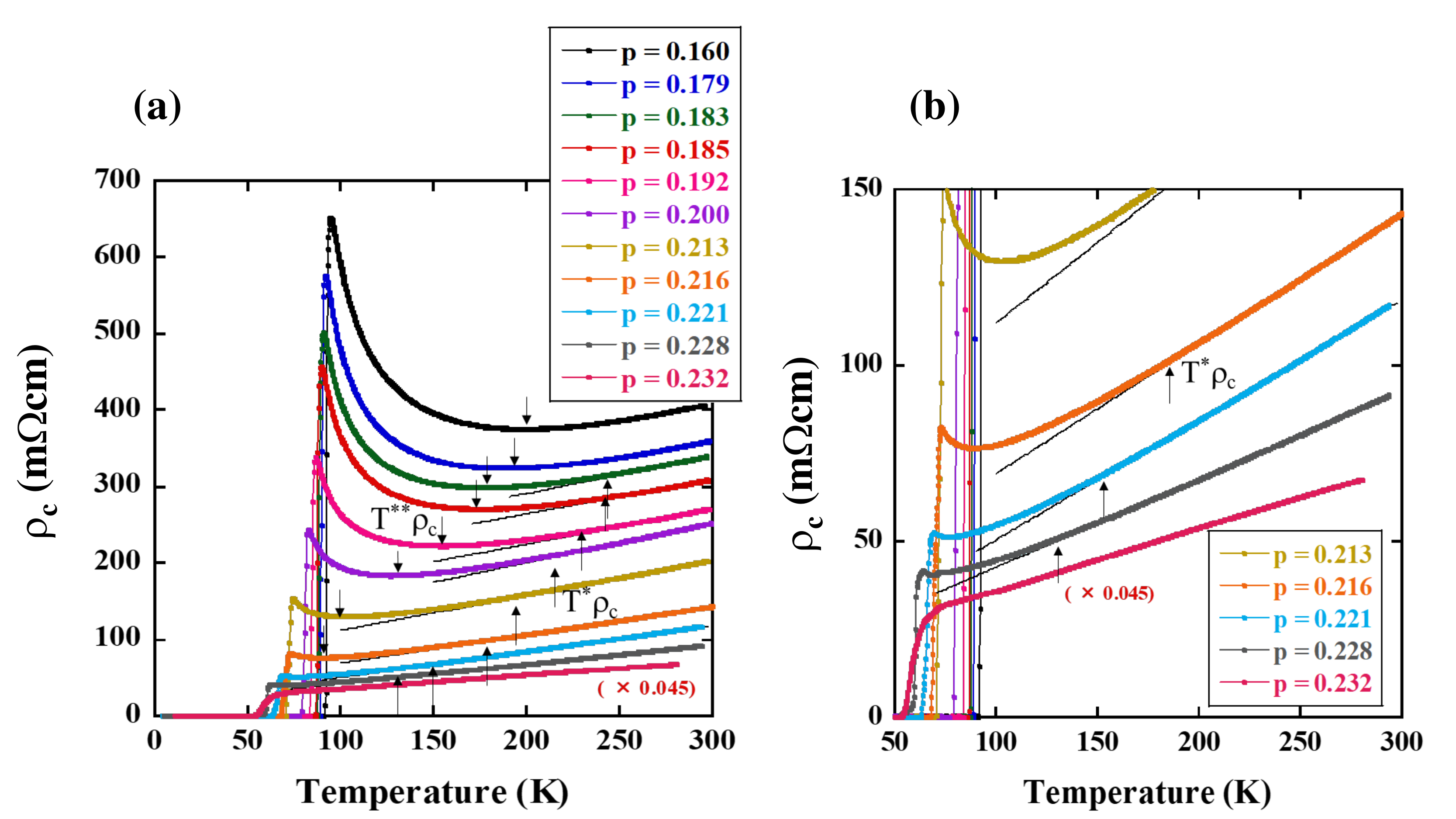}
			\caption{\label{fig4}(Color online)(a)Out-of-plane resistivity $\rho_{c}(T)$ for Bi$_{1.6}$Pb$_{0.4}$Sr$_2$CaCu$_2$O$_{8+\delta}$ single crystals with various $p$. The most overdoped sample with $p$ = 0.232 is pristine Bi$_{2}$Sr$_2$CaCu$_2$O$_{8+\delta}$ \cite{Usui14}. The temperatures $T^{*}_{\rho_{c}}$ below which $\rho_{c}(T)$ gradually increases and $T^{**}_{\rho_{c}}$ at which $\rho_{c}(T)$ is minimized are indicated by arrows. (b) Expanded view of the overdoped side of (a).}
		\end{center}
	\end{figure}

Figure \ref{fig4}(a) shows the temperature dependence of the out-of-plane resistivity, $\rho_{c} (T)$ for Bi$_{1.6}$Pb$_{0.4}$Sr$_2$CaCu$_2$O$_{8+\delta}$ single crystals with various $p$. The most overdoped sample ($p$ = 0.232) is pristine Bi$_{2}$Sr$_2$CaCu$_2$O$_{8+\delta}$. Although $\rho_{c} (T)$ exhibits metallic behavior above 250 K, the gradual upturn below a certain temperature $T^*_{\rho_{c}}$ is an indication of the opening of the pseudogap \cite{Watanabe97,Shibauchi01,Watanabe00,Usui14} (hereafter referred to as ``weak'' pseudogap \cite{Norman05}). $T^*_{\rho_{c}}$ is determined using a previously reported definition \cite{Watanabe97,Watanabe00,Usui14}. $T^*_{\rho_{c}}$ is high even for $p$ $>$ 0.19 but decreases rapidly when $p$ exceeds 0.22. However, $T^*_{\rho_{c}}$ cannot be determined above $T_c$ at $p$ = 0.232 (Fig. \ref{fig4}(b)). At temperatures below $T^*_{\rho_{c}}$, $\rho_{c} (T)$ increases rapidly below $T^{**}_{\rho_{c}}$. This suggests that another pseudogap (hereafter referred to as ``strong'' pseudogap \cite{Norman05}) opens up below this temperature. $T^{**}_{\rho_{c}}$ is estimated as 198, 189, 178, 176, 153, 129, 101, and 90 K for $p$ = 0.160, 0.179, 0.183, 0.185, 0.192, 0.200, 0.213, and 0.216, respectively. Here, $T^{**}_{\rho_{c}}$ is defined as the temperature at which $\rho_{c} (T)$ is minimized. $T^{**}_{\rho_{c}}$ cannot be determined for $p$ $\ge$ 0.221, because the rapid increase is not observed in these samples.

\begin{figure}[t]
		\begin{center}
			\includegraphics[width=85mm]{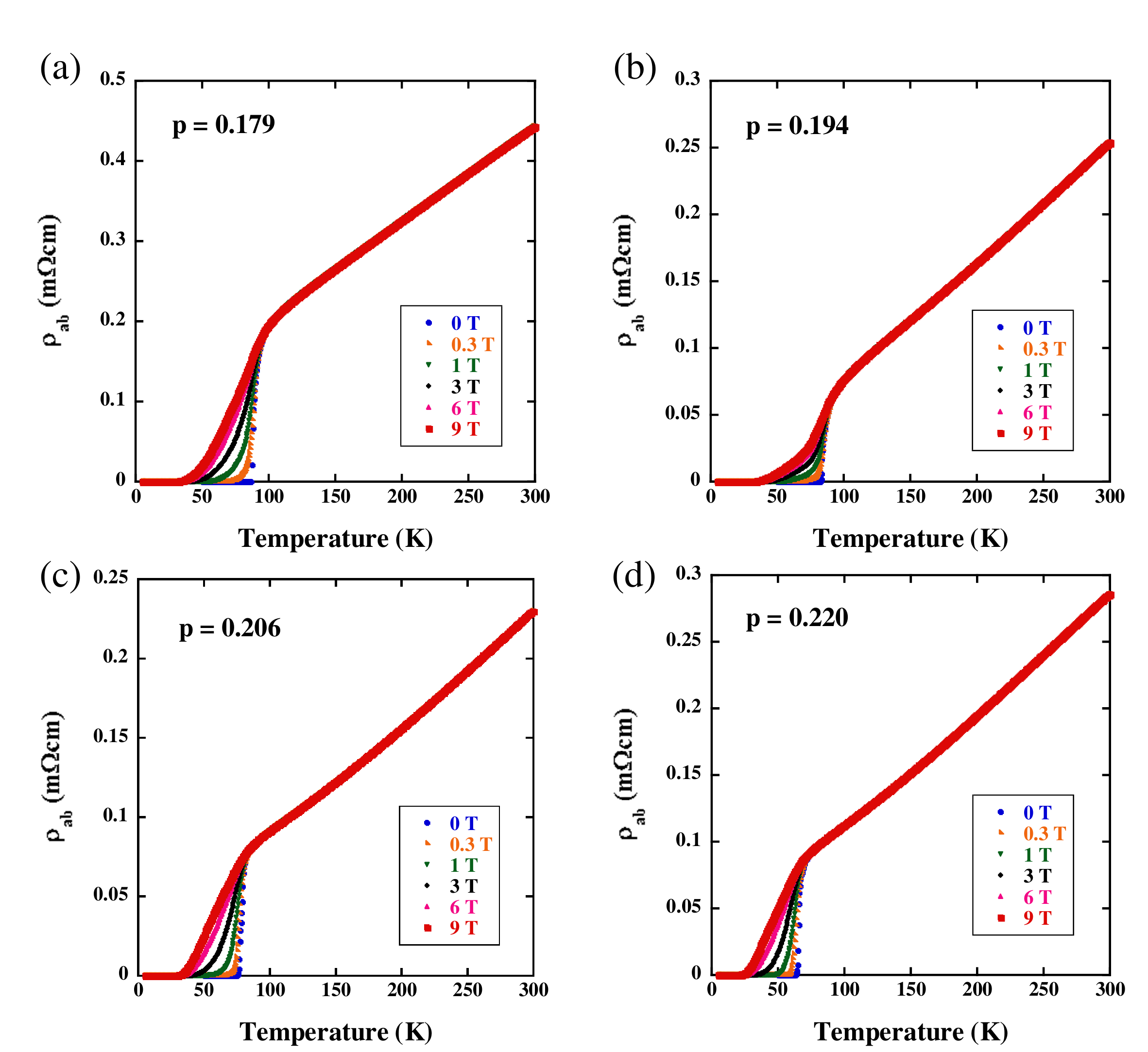}
			\caption{\label{fig5}(Color online) In-plane resistivity $\rho_{ab}(T)$ for Bi$_{1.6}$Pb$_{0.4}$Sr$_2$CaCu$_2$O$_{8+\delta}$ single crystals under various magnetic fields $B \parallel c$ with (a) $p$ = 0.179, (b) $p$ = 0.194, (c) $p$ = 0.206, and (d) $p$ = 0.220, respectively.}
		\end{center}
	\end{figure}

\begin{figure*}[t]
		\begin{center}
			\includegraphics[width=150mm]{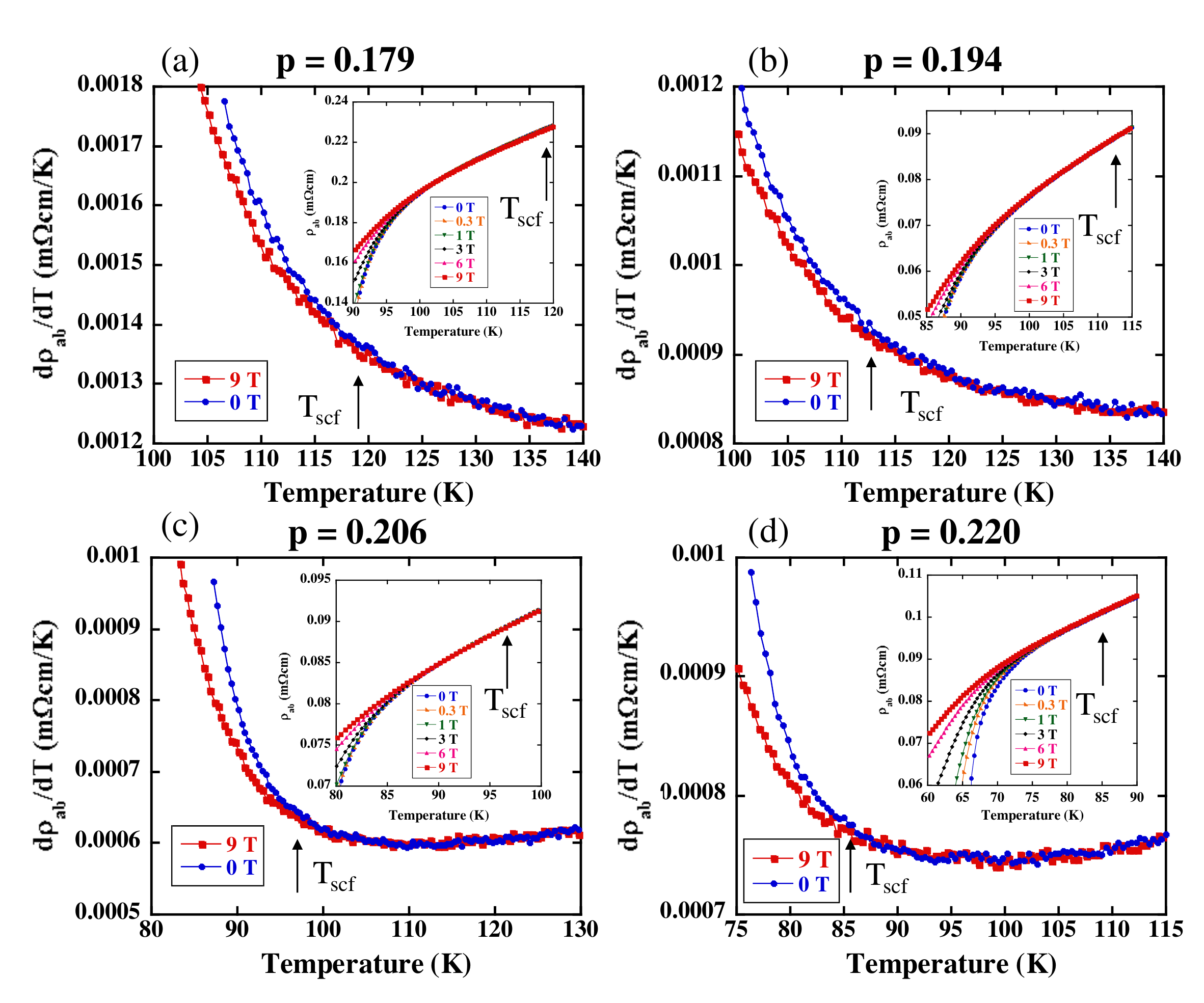}
			\caption{(Color online)
				Temperature derivative of $\rho_{ab}$ vs. the temperatures at 0 T and 9 T for (a) $p$ = 0.179, (b) $p$ = 0.194, (c) $p$ = 0.206, and (d) $p$ = 0.220. The temperatures $T_{scf}$ below which $d\rho_{ab}/dT$ at 9 T deviates from that at 0 T are indicated by arrows. The inset shows an expanded plot of $\rho_{ab}$ vs. temperature near $T_c$ under various magnetic fields up to 9 T. }
\label{fig6}
			
		\end{center}
	\end{figure*}

To investigate the relationship between the ``weak'' and ``strong'' pseudogap opening temperatures and $T_{scf}$, $T_{scf}$ was estimated by comparing the values of $\rho_{ab} (T)$ with and without a magnetic field. Figure \ref{fig5}(a), \ref{fig5}(b), \ref{fig5}(c), and \ref{fig5}(d) show $\rho_{ab} (T)$ under various magnetic fields up to 9 T for the samples with $p$ = 0.179, 0.194, 0.206, and 0.220, respectively. The temperature dependence of the normal state reproduce the features shown in Fig. \ref{fig2} well. In the superconducting state, a typical fan-shaped broadening under the magnetic fields due to the suppression of the Aslamazov--Larkin type superconducting fluctuation effect in strongly two-dimensional superconductors \cite{Ikeda91,Livanov00,Adachi15} was observed. The expanded plots near $T_c$ are shown in the insets to Figs. \ref{fig6}(a)--(d). Since the largest field, 9 T, was most effective in suppressing the superconducting fluctuation effect, we depict the temperature derivative of $\rho_{ab} (T)$ at 0 T and 9 T as a function of the temperature for each doping level in Figs. \ref{fig6}(a)--(d). In all figures, the data plots at 9 T deviate from those at 0 T below a certain temperature, $T_{scf}$, indicating that the superconducting fluctuation effect manifests below these temperatures. Based on this, $T_{scf}$ is defined as the temperature at which $d\rho_{ab} (T)/dT$ under a magnetic field of 9 T decreases by 1 $\%$, relative to that at 0 T. Consequently, $T_{scf}$ is estimated as 119, 113, 97, and 86 K for $p$ = 0.179, 0.194, 0.206, and 0.220, respectively. This result is consistent with our previous study on Bi-2212 \cite{Usui14}. $T_{scf}$ for $p$ = 0.232 was previously estimated as 73 K \cite{Usui14}. 

\begin{figure}[t]
		\begin{center}
			\includegraphics[width=85mm]{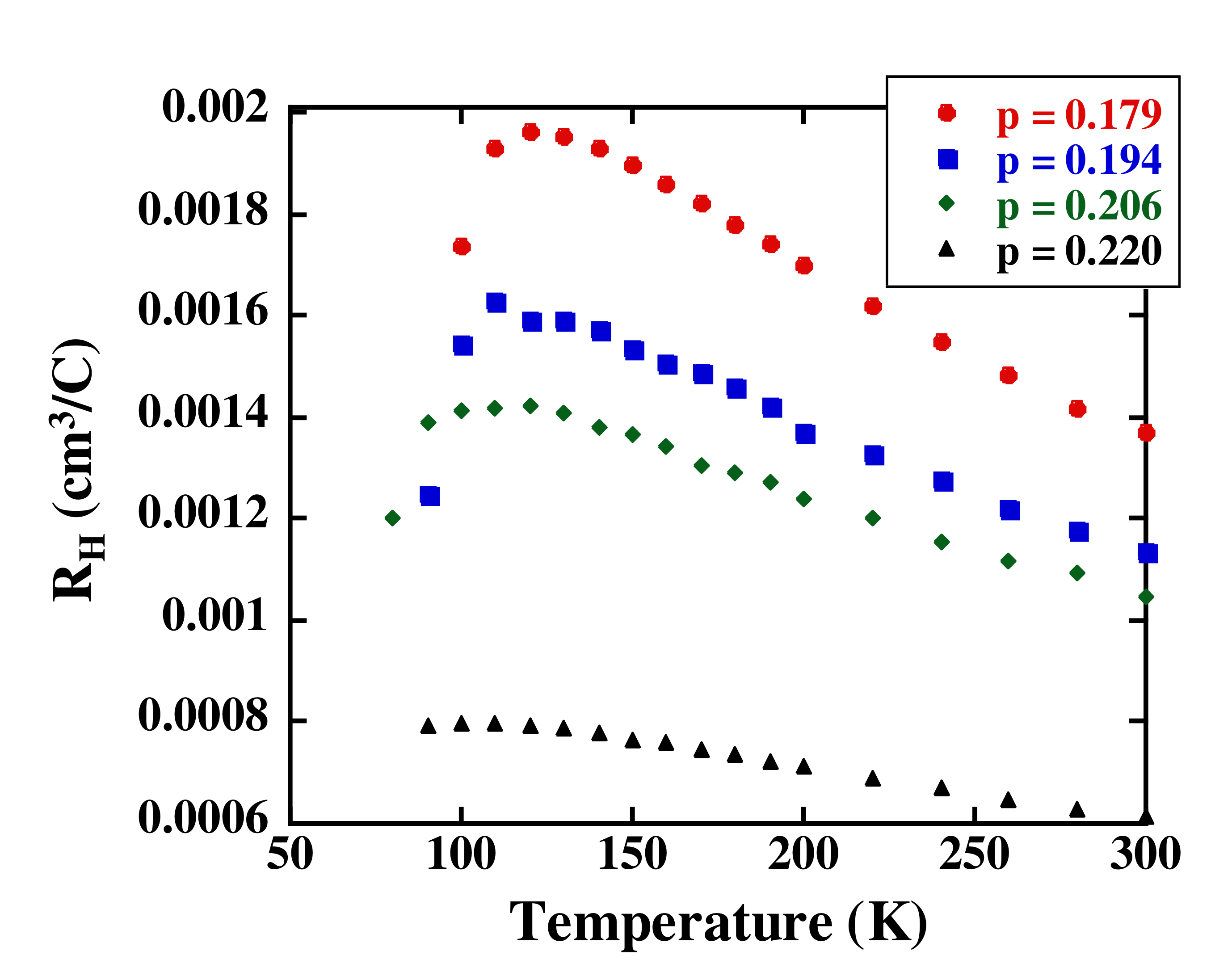}
			\caption{\label{fig7}(Color online) Hall coefficient $R_{H} (T)$ for Bi$_{1.6}$Pb$_{0.4}$Sr$_2$CaCu$_2$O$_{8+\delta}$ single crystals with various $p$. The red solid circles, blue solid squares, green solid diamonds, and black solid up-pointing triangles represent data points for $p$ = 0.179, 0.194, 0.206, and 0.220, respectively.}
			
		\end{center}
	\end{figure}

\begin{figure}[t]
		\begin{center}
			\includegraphics[width=85mm]{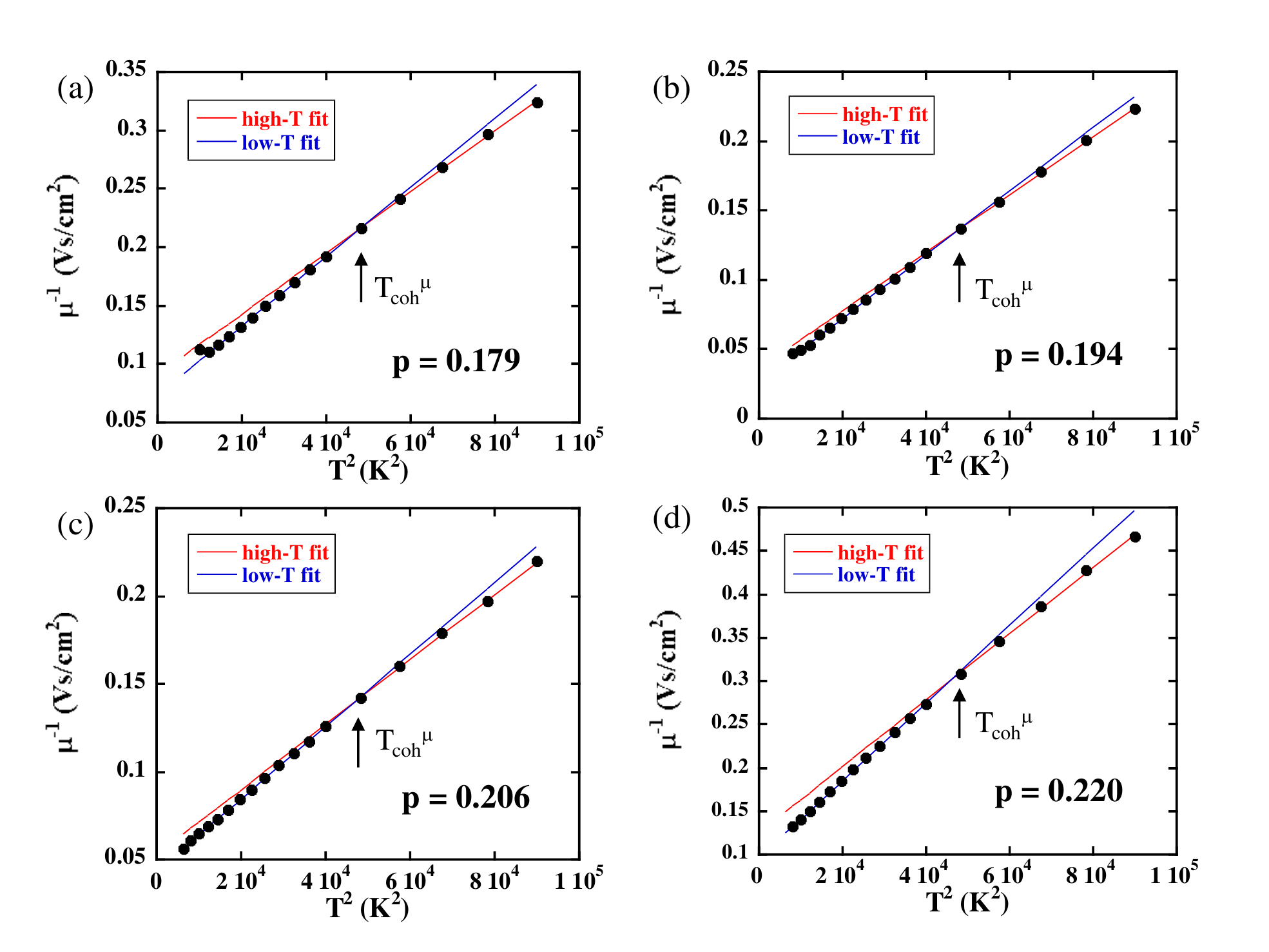}
			\caption{\label{fig8}(Color online) Inverse Hall mobility $\mu^{-1}$ vs. $T^2$ for Bi$_{1.6}$Pb$_{0.4}$Sr$_2$CaCu$_2$O$_{8+\delta}$ single crystals with (a) $p$ = 0.179, (b) $p$ = 0.194, (c) $p$ = 0.206, and (d) $p$ = 0.220. The red straight lines denote high-temperature linear fits between 220 and 300 K. The blue straight lines denote low-temperature linear fits between (a) 120 and 190 K, (b) 110 and 190 K, (c) 110 and 190 K, and (d) 90 and 190 K, respectively. $T_{coh}{\mu}$ at which the high-temperature and low-temperature linear fits intersect are indicated by arrows.}
			
		\end{center}
	\end{figure}

To obtain further insight into the occurrence of incoherent to coherent crossover transition, we measured $R_{H} (T)$ for the same samples shown in Figs. \ref{fig5}(a)--(d). The results are shown in Fig. \ref{fig7}. The $\rho_{yx}$ was positive and linear in $B$ within the temperature range measured. The magnitude of $R_{H} (T)$ decreases with increasing $p$, ensuring the increase in hole concentration with $p$. The temperature dependence of $R_{H} (T)$ decreases with increasing $p$, which agrees with previous studies on LSCO \cite{Ando04}, Tl-2201 \cite{Kubo91}, and BSLCO \cite{Ando99}. Then, combining the results of Fig. \ref{fig5} and Fig. \ref{fig7}, Hall mobility $\mu (T)$ is estimated. Figure \ref{fig8}(a), \ref{fig8}(b), \ref{fig8}(c), and \ref{fig8}(d) show $\mu^{-1} (T)$ as a function of $T^2$ for the samples with $p$ = 0.179, 0.194, 0.206, and 0.220, respectively. In all figures, $\mu^{-1} (T)$ roughly obeys the empirical $\propto T^2$ relation \cite{Anderson91,Ong91,Kontani08,Barisic19} at higher temperatures above 200 K, whereas it slightly deviates downward below  200 K, indicating that, below 200 K, $\mu$ is larger than the values that extrapolated high temperature values to low temperatures using the relation, $\mu \propto T^{-2}$. This implies that the electronic system acquires coherence below 200 K. Based on this observation, $T_{coh}\mu$ is defined as the temperature at which linearly fitted lines above and below 200 K intersect. Consequently, $T_{coh}\mu$ is estimated as 221, 217, 215, and 214 K for $p$ = 0.179, 0.194, 0.206, and 0.220, respectively. It should be noted that similar deviation from high-temperature $\propto T^2$ behaviors at lower temperatures have been reported for the Hall angle, cot$\theta_H$ (a quantity similar to $\mu^{-1}$) of overdoped cuprates \cite{Ando99, Ando04}, although the authors of refs. \cite{Ando99} and \cite{Ando04} did not interpret this deviation as evidence for electronic coherence.

\section{PHASE DIAGRAM}
\begin{figure}[t]
		\begin{center}
			\includegraphics[width=85mm]{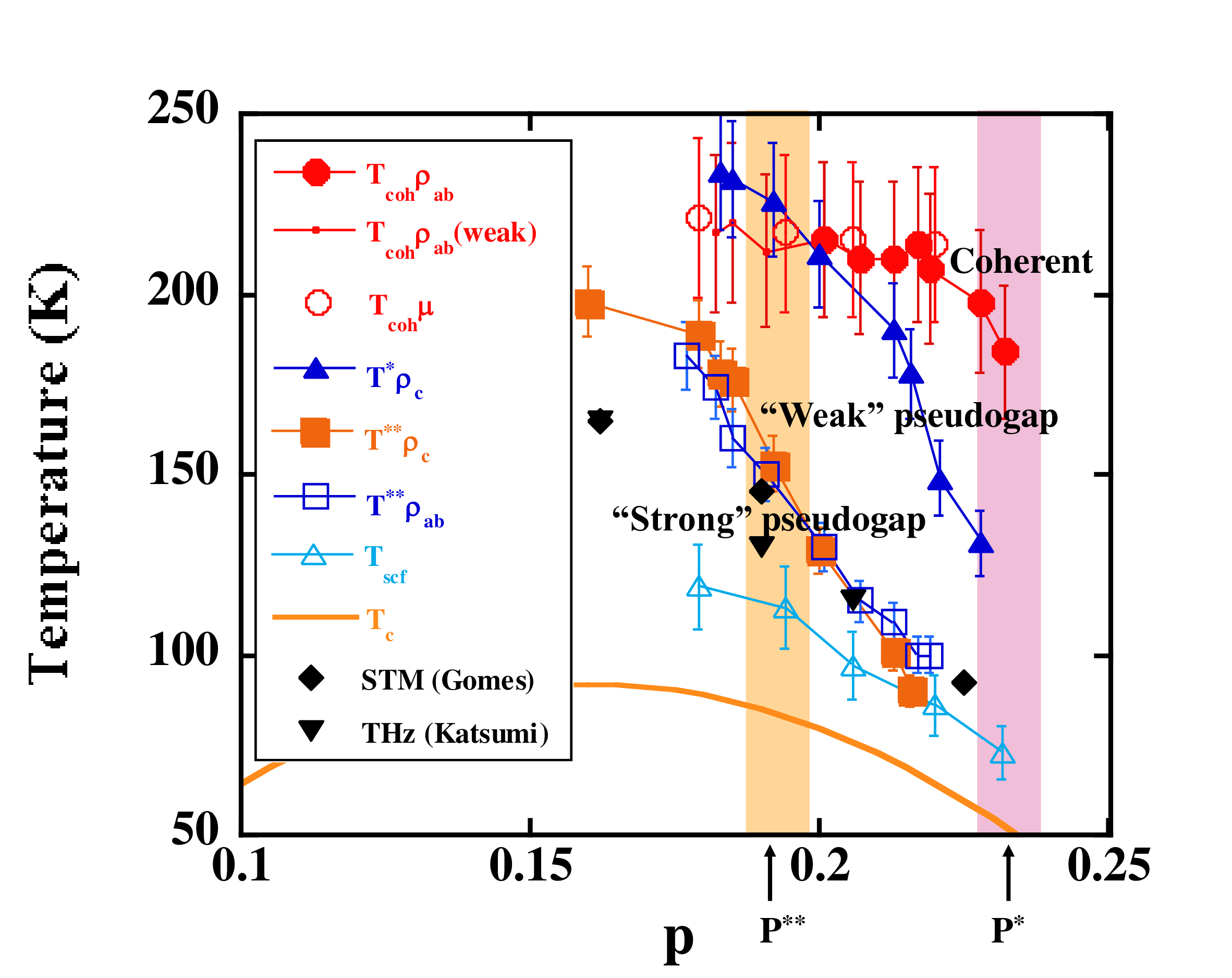}
			\caption{\label{fig9}(Color online) Characteristic temperatures vs. $p$ for Bi$_{1.6}$Pb$_{0.4}$Sr$_2$CaCu$_2$O$_{8+\delta}$ and Bi$_{2}$Sr$_2$CaCu$_2$O$_{8+\delta}$ \cite{Usui14} single crystals. The black closed diamonds represent the pseudogap opening temperatures obtained by scanning tunneling microscopy (STM) \cite{Gomes07} and down-pointing triangles represent those obtained by terahertz (THz) pump-optical probe spectroscopy  \cite{Katsumi20}. The orange and purple bands indicate the characteristic doping levels of $p^{**}$ $\approx$ 0.19 and $p^{*}$ $\approx$ 0.23, respectively.}
			
		\end{center}
	\end{figure}

The characteristic temperatures $T_{coh}$, $T^*_{\rho_{c}}$, $T^{**}_{\rho_{c}}$, $T^{**}_{\rho_{ab}}$, and $T_{scf}$ are plotted as functions of $p$ in Fig. \ref{fig9}. The plots show that $T_{coh}{\rho_{ab}}$ and $T_{coh}\mu$ coincide, which indicates that both have the same origin (i.e., from incoherent to coherent crossover temperatures). Based on this, we interpret $T_{coh}$ as representing $T_{coh}{\rho_{ab}}$ and $T_{coh}\mu$ collectively. The plots also show that $T^{**}_{\rho_{c}}$ and $T^{**}_{\rho_{ab}}$ coincide, which indicates again that both have the same origin (i.e., the ``strong'' pseudogap opening). The rapid decrease in $\rho_{ab}(T)$ below $T^{**}_{\rho_{ab}}$ may be due to the decreasing scattering rate upon opening of the ``strong'' pseudogap. On this basis, we denote $T^{**}$ as representing $T^{**}_{\rho_{c}}$ and $T^{**}_{\rho_{ab}}$ collectively. For $p$ $\le$ 0.19, $T^{**}$ differs from $T_{scf}$, whereas for $p$ $>$ 0.19, $T^{**}$ rapidly approaches $T_{scf}$, and eventually coincides with it at $p$ $\approx$ 0.21. 

Next, we discuss the end point, $p^*$, of the ``weak'' pseudogap. Figure \ref{fig9} shows that $T^{*}_{\rho_{c}}$ decreases rapidly for $p$ $>$ 0.22, and then disappears at $p$ $\approx$ 0.23. This result suggests that $p^*$ $\approx$ 0.23 for the Bi-2212 case. As discussed in Fig. \ref{fig3}(c), we observed a kink in the $T^2$ coefficient, $\alpha_{2}$, of $\rho_{ab} (T)$ at $p$ = 0.22, where Lifshitz transition occurs \cite{Kaminski06}. Then, the disappearance for $T^{*}_{\rho_{c}}$ at $p$ $\approx$ 0.23 may be attributed to this Lifshitz transition, since the spectral weights at anti-nodal directions, which are needed to open the ``weak'' pseudogap, are expected to decrease after the transition \cite{Benhabib15}.

Finally, in this section, we demonstrate the positional relationship between $T_{coh}$ and $T^{*}_{\rho_{c}}$. Figure \ref{fig9} shows that $T_{coh}$ is independent of the doping level, with a slight tendency to decrease with excessive overdoping. $T_{coh}$ seems to exist even below $p$ = 0.19, since $T_{coh}\mu$ can be defined at $p$ = 0.179 (Fig. \ref{fig8}(a)), although the change in $d\rho_{ab}/{dT}$ above and below $T_{coh}{\rho_{ab}}$ is insignificant for $p$ $<$ 0.19 (see Supplemental Materials \cite{Supplemental}). This insignificance may be attributed to the fact that the feature accompanied by electronic coherence is masked by the incipient ``strong'' pseudogap opening. In fact, $T_{coh}{\rho_{ab}}$ can be seen more clearly in BSLCO for 0.16 $\le$ $p$ $\le$ 0.18 \cite{Ando04_1} (In Fig. 1(b) of ref. \cite{Ando04_1}, with decreasing temperatures across $T_{coh}{\rho_{ab}}$ $\approx$ 200 K, the thin red colored region evolves into the thick colored region representing that $d^2\rho_{ab}/{dT^2}$ is enhanced upon cooling). This may be due to the fact that in BSLCO, $T^{**}$ is lower ($<$ 100 K) than that in Bi(Pb)-2212. (In Fig. 1(b) of ref. \cite{Ando04_1}, $T^{**}$ in our definition is depicted as the white band between the red and blue region for 0.16 $\le$ $p$ $\le$ 0.18). Thus, the feature accompanied by electronic coherence was not masked in BSLCO. We note strong correlations of these observations for BSLCO with those of cot$\theta_H (T)$. This means that in BSLCO, cot$\theta_H (T)$ shows marked deviation from the empirical $\propto T^2$ relation in the overdoped state \cite{Ando99}, which agrees with our identification of $T_{coh}$. Based on these results, we conclude that $T^{*}_{\rho_{c}}$ and $T_{coh}$ intersect at $p$ = 0.19.

\section{DISCUSSION}
Here, we discuss the implications of the obtained results. First, we accurately determined the characteristic temperatures of two types of pseudogaps ($T^{*}_{\rho_{c}}$ and $T^{**}$ for the ``weak'' and ``strong'' pseudogaps, respectively) by combining results from in-plane and out-of-plane transport measurements (Fig. \ref{fig9}). Then, we showed that $T^{**}$ differed for $T_{scf}$ for $p$ $\le$ 0.19, while they coincided for $p$ $>$ 0.19. Based on this result, we considered that the ``strong'' pseudogap originated from preformed Cooper pairing (phase-incoherent Cooper pair formation) in the QCP model (Fig. \ref{fig1}(a)) or from spinon pairing in the RVB model (Fig. \ref{fig1}(b)). Although several studies have reported preformed Cooper pairing in high-$T_c$ cuprates \cite{Kondo13,Reber12,Reber13,Kondo11,Kondo15,Gomes07,Dubroka11,Uykur14,Lee17,Wang06,Katsumi20,Li10}, a general consensus has not been realized \cite{Keimer15}. Furthermore, the doping levels for which these distinct pseudogaps terminate are unknown. We showed that $T^{**}$ ends (or merges with $T_{scf}$) at approximately $p^{**}$ = 0.19, whereas $T^{*}_{\rho_{c}}$ ends at approximately $p^{*}$ = 0.23 for Pb-doped Bi-2212. Given that $p^{*}$ of YBa$_2$Cu$_3$O$_{7-\delta}$ is 0.19 \cite{Tallon01}, it may be material (i.e., band structure)-dependent \cite{Benhabib15}.

Next, we found that $T^{*}_{\rho_{c}}$ and $T_{coh}$ cross each other at $p^{**}$ = 0.19. This behavior is incompatible with the conventional QCP scenario (Fig. \ref{fig1}(a)) \cite{Varma99,Sachdev10}. Instead, our results favor the RVB scenario (Fig. \ref{fig1}(b)). In the QCP scenario, $T^{*}$ represents the temperatures of the phase transition \cite{Varma99,Sachdev10}, whereas in the RVB model, $T_{D}$($T_{D}^{(0)}$) represents crossover temperatures from the strange metal to the pseudogap (spin gap to be precise) states \cite{Lee06, Ogata08, Lee92}. In recent years, several phase transitions with symmetry breaking have been observed using various techniques in the pseudogap regime \cite{Tranquada95,Ghiringhelli12,Hanaguri04,Fauque06,Xia08,Hashimoto10,Ando02,Daou10,Ishida20,Sato18}, which may support the QCP scenario. We encountered a serious problem on reconciling these observations with our results, which is currently an open question. In this regard, Hussey argued that such phase transitions do not form a pseudogap, but instead involve the development of electronic instability inside the pseudogap regime \cite{Hussey18}. Tallon and Loram argued that the pseudogap reflects an underlying energy scale $E_{g}$ that vanishes beneath the superconducting dome \cite{Tallon20}. The energy scale may correspond to the superexchange energy, $J$, and provide the crossover temperatures. Thus, we assumed that $T^{*}_{\rho_{c}}$ corresponds to $T_{D}^{(0)}$ of the RVB model \cite{Lee06, Ogata08, Lee92}.

The ``strong'' pseudogap effect has been predicted using theories based on the Fermi-liquid point of view, which considers the effect of large antiferromagnetic spin fluctuations in quasi-two-dimensional metals near the Mott insulator \cite{Kobayashi01,Yanase01,Onoda00}. In the RVB model, the ``strong'' pseudogap effect may be attributed to spinon pairing below $T_{D}$ \cite{Lee06, Ogata08, Lee92}. Furthermore, $T_{coh}$ may correspond to the temperatures for Bose condensation of holons, $T_{BE}$, in the RVB model \cite{Lee92}. In this model, $T_{BE}$ is proportional to the superfluid density $\rho_{s} (0)$. Our result of approximately doping-independent $T_{coh}$ in the overdoped region agrees with experimental observation for $\rho_{s} (0)$ \cite{Tallon03}. Consequently, our observed phase diagram (Fig. \ref{fig9}) roughly coincides with the prediction of the RVB model (Fig. \ref{fig1}(b)).

It is intriguing to note that this normal state crossing phenomena of $T^{*}_{\rho_{c}}$ and $T_{coh}$ influences the nature of superconducting fluctuations. When $T^{*}_{\rho_{c}}$ $<$ $T_{coh}$ for $p$ $>$ 0.19, the phase ordering of Cooper pairs starts to develop almost simultaneously as they form below $T^{**}$ $\approx$ $T_{scf}$. When $T^{*}_{\rho_{c}}$ $>$ $T_{coh}$ for $p$ $<$ 0.19, Cooper pairs are preformed below $T^{**}$, but their phases are not settled. Thus, superconducting fluctuation does not appear until lower values of $T_{scf}$. This non-superconducting behavior between $T_{scf}$ and $T^{**}$ when $p$ $<$ 0.19 is consistent with the spinon pairing in the RVB model \cite{Lee06, Ogata08, Lee92}. In addition, with decreasing $p$ below 0.19, $T^{*}_{\rho_{c}}$ and $T_c$ increase. This result implies that the pseudogap does not just compete with superconductivity, but reflects the energy scale, $J$, which is probably the source for superconductivity.

Although our data are incompatible with the conventional QCP scenario, $p^{**}$ = 0.19 may be an anomalous critical point. In this study, we found that $T^{*}_{\rho_{c}}$ and $T_{coh}$ cross at $p^{**}$, and the preformed nature for Cooper pairing changes over this doping level. Therefore, $p^{**}$ = 0.19 is closely related to superconductivity and may be universal for hole-doped cuprates \cite{Tallon20,Hussey18}. This anomalous criticality at $p^{**}$ was first proposed by Tallon and Loram \cite{Tallon01}.

Here, we assumed Eq. (1) and the incoherent to coherent crossover at $T_{coh}$ $\approx$ 200 K to interpret the temperature dependence of $\rho_{ab} (T)$. Alternatively, the power law formula may also reproduce the observed $\rho_{ab} (T)$ \cite{Dessau19}. Another interpretation may be the achievement of the quantum mechanical constraint for the maximum scattering rate (Planckian dissipation limit) \cite{Zaanen04,Hussey11,Hussey18}.

\section{SUMMARY}
In this study, we clarified the true phase diagram for the overdoped side of high-$T_c$ cuprates. The frequently discussed critical doping level, $p^{**}$ = 0.19, is a doping level above which $T_{coh}$ can be observed clearly, and the characteristic temperature of the opening of the ``strong'' pseudogap, $T^{**}$, approaches $T_{scf}$ rapidly. These results agree with the RVB scenario, rather than the conventional QCP scenario. However, an anomalous QCP scenario cannot be ruled out. To complete the phase diagram, our approach must be extended to the underdoped side.

\section*{Acknowledgments}
The authors would like to acknowledge useful discussions with Y. Koike, T. Kondo, T. Shibauchi, and M. Ogata. This work was supported by a Hirosaki University Grant for Distinguished Researchers FY2017-2018. The magnetotransport measurements were performed using PPMS at Iwate University, the Institute for Solid State Physics, University of Tokyo, and Cryogenic Research Center, University of Tokyo. This work was supported by JSPS KAKENHI Grant Number 20K03849.




\end{document}


\preprint{APS/123-QED}

\title{Revised phase diagram of the high-$T_c$ cuprate superconductor Pb-doped Bi$_2$Sr$_2$CaCu$_2$O$_{8+\delta}$ revealed by anisotropic transport measurements} 

\author{Keiichi Harada}
\affiliation{Graduate School of Science and Technology, Hirosaki University, Hirosaki, Aomori, 036-8561 Japan}
	\author{Yuki Teramoto}
\affiliation{Graduate School of Science and Technology, Hirosaki University, Hirosaki, Aomori, 036-8561 Japan}
	\author{Tomohiro Usui}
\affiliation{Graduate School of Science and Technology, Hirosaki University, Hirosaki, Aomori, 036-8561 Japan}
	\author{Kenji Itaka}
 \affiliation{Institute of Regional Innovation, Hirosaki University, Aomori, Aomori, 036-8561 Japan}
        \author{Takenori Fujii}
\affiliation{Cryogenic Research Center, University of Tokyo, Bunkyo, Tokyo 113-0032, Japan}
        \author{Takashi Noji}
 \affiliation{Graduate School of Engineering, Tohoku University, Sendai 980-8579, Japan}
	\author{Haruka Taniguchi}
\affiliation{Graduate School of Engineering, Iwate University, Morioka 020-8551, Japan}
	\author{Michiaki Matsukawa}
\affiliation{Graduate School of Engineering, Iwate University, Morioka 020-8551, Japan}
	\author{Hajime Ishikawa}
\affiliation{Institute for Solid State Physics, University of Tokyo, Kashiwa, Chiba 277-8581, Japan}
        \author{Koichi Kindo}
\affiliation{Institute for Solid State Physics, University of Tokyo, Kashiwa, Chiba 277-8581, Japan}
        \author{Daniel S. Dessau}
 \affiliation{Department of Physics, University of Colorado at Boulder, Boulder, CO 80309, USA}

	\author{Takao Watanabe}
\email{twatana@hirosaki-u.ac.jp}
\affiliation{Graduate School of Science and Technology, Hirosaki University, Hirosaki, Aomori, 036-8561 Japan}





\date{\today}

\maketitle

\section*{A. Sample preparation}

To control the doping level ($p$), single crystals were annealed under the conditions listed in Table \ref{tab1}. After annealing, they were rapidly cooled to room temperature at a rate of approximately 30 $^\circ$ C/min under the same conditions. 

\begin{table}

   \caption{Summary of annealing conditions, $T_c$, and their doping levels ($p$) used in the in-plane resistivity measurements. The same conditions were used for out-of-plane resistivity measurements, except for slight changes in $T_c$ and $p$.}
   \label{tab1}
   \vspace*{0.5cm}

\begin{tabular}{ccc}
\hline
\hline
Annealing Conditions & $T_c$ & $p$ \\
\hline
600 $^\circ$C, 24 h in O$_2$ 2 Pa  & 89.6 K & 0.177 \\
600 $^\circ$C, 12 h in O$_2$ 10 Pa  & 87.0 K & 0.185 \\
600 $^\circ$C, 12 h in O$_2$ 100 Pa  & 84.5 K & 0.191 \\
600 $^\circ$C, 18 h in O$_2$ 1 $\%$  & 79.0 K & 0.201 \\
600 $^\circ$C, 18 h in O$_2$ 10 $\%$  & 75.0 K & 0.207 \\
600 $^\circ$C, 36 h in O$_2$ 100 $\%$  & 70.7 K & 0.213 \\
500 $^\circ$C, 70 h in O$_2$ 100 $\%$  & 67.4 K & 0.217 \\
400 $^\circ$C, 200 h in O$_2$ 100 $\%$  & 65.1 K & 0.219 \\
400 $^\circ$C, 200 h in O$_2$ 100 $\%$ 420 atm & 56.3 K & 0.228 \\
400 $^\circ$C, 30 h in O$_2$ 100 $\%$ 400 atm (pristine) & 52.0 K & 0.232 \\
\hline
\hline
\end{tabular} 
\end{table}

\section*{B. Information on the linear fits for the temperature derivative of in-plane resistivity}

\begin{figure*}[t]
		\begin{center}
			
			\includegraphics[width=150mm]{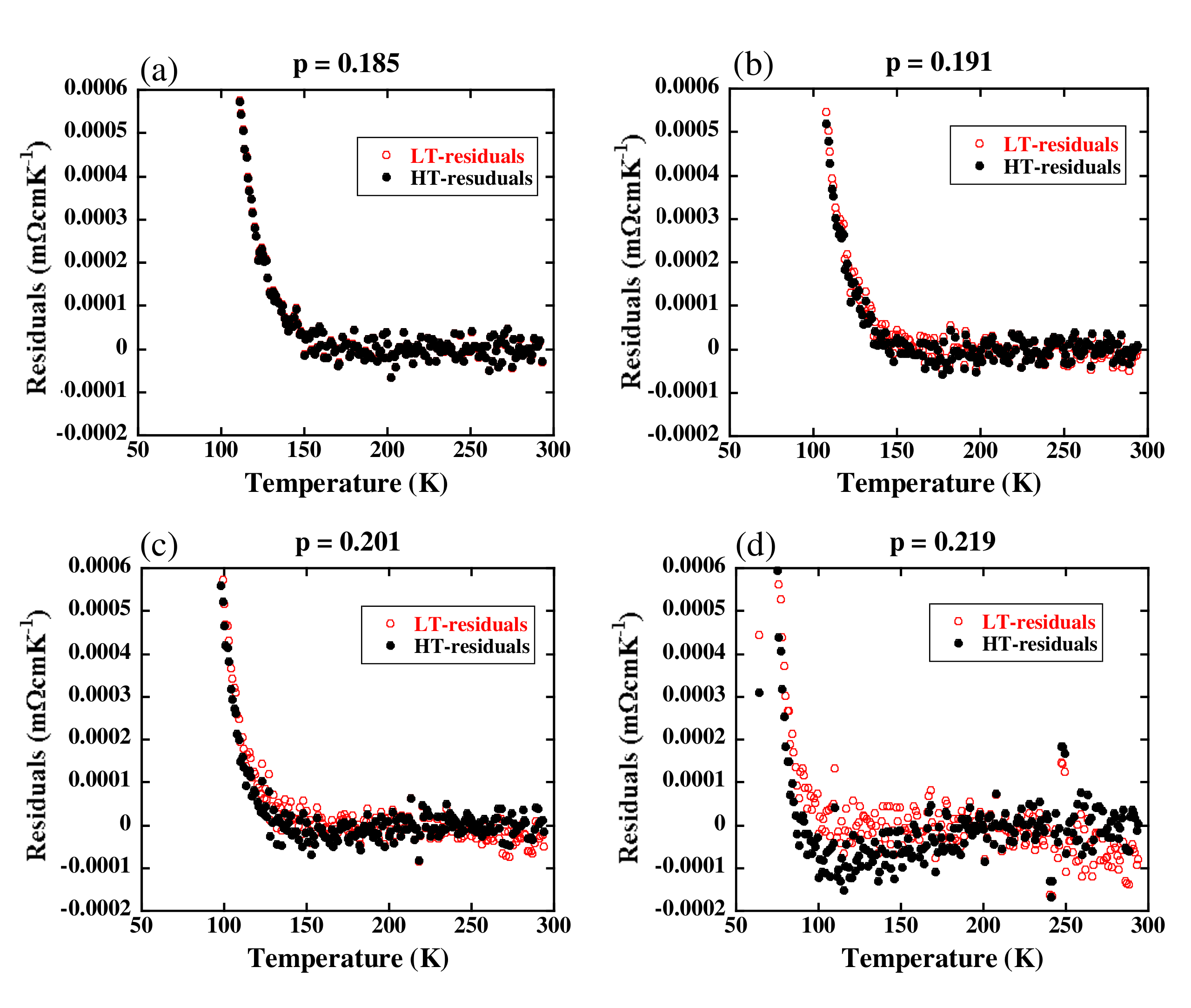}
			\caption{(Color online)
				Temperature dependence of residuals after subtracting linear fits from the data points of $d\rho_{ab} (T)/dT$ for (a) $p$ = 0.185, (b) $p$ = 0.191, (c) $p$ = 0.201, and (d) $p$ = 0.219 respectively. In each figure, black solid circles and red open circles represent residuals after subtracting high-temperature and low-temperature linear fits, respectively.}
\label{figS1}
			
		\end{center}
	\end{figure*}

Here, we provide more information on linear fits for the temperature derivative of in-plane resistivity $\rho_{ab} (T)$ (Fig. 3(a) in the main text). The high-temperature linear fits for $d\rho_{ab} (T)/dT$ were performed between 220 and 300 K for all the samples. The low-temperature linear fits for $d\rho_{ab} (T)/dT$ were performed between 170 and 190 K, 160 and 190 K, 150 and 190 K, and 140 and 190 K for $p$ = 0.185, 0.191, 0.201, and 0.219, respectively. Figure \ref{figS1}(a), \ref{figS1}(b), \ref{figS1}(c), and \ref{figS1}(d) show each residual after subtracting both high- and low-temperature linear fits from the data points for $p$ = 0.185, 0.191, 0.201, and 0.219, respectively. When the doping level $p$ = 0.185, the difference between high- and low-temperature linear fits is insignificant. This may be attributed to the narrow fitting temperature interval, which is caused by the incipient ``strong'' pseudogap opening at lower temperatures. The fact that the change between high- and low-temperature linear fits is insignificant is stated in the former half of the fourth paragraph of section III in the main text. In addition, the possible reason is discussed in more detail in the last paragraph of section IV in the main text. On the other hand, when increasing $p$ above 0.19, the difference between high- and low-temperature linear fits become prominent, indicating that the system crossovers from the high-temperature strange metal state to the low-temperature Fermi-liquid-like state more clearly with the increasing doping level.

